\documentclass[journal,draftclsnofoot,onecolumn,12pt,oneside]{IEEEtran} 

\usepackage{cite}

\ifCLASSINFOpdf
  \usepackage[pdftex]{graphicx}
  \usepackage{epstopdf} 
\else
\fi
\usepackage{amsmath}
  \usepackage[caption=false,font=footnotesize]{subfig}
\usepackage{url}

\newcommand{\comment}[1]{{}}

\usepackage{amssymb}
\usepackage{xspace}

\usepackage{bm}

\newcommand{\ctrans}{\ensuremath{^{{*}}}\xspace}

\newcommand{\mat}[1]{\ensuremath{\mathbf{#1}}\xspace} 

\newcommand{\prebb}{\ensuremath{\mat{F}_{\mathrm{BB}}}\xspace} 

\newcommand{\prerf}{\ensuremath{\mat{F}_{\mathrm{RF}}}\xspace} 
\newcommand{\combb}{\ensuremath{\mat{W}_{\mathrm{BB}}}\xspace} 
\newcommand{\comrf}{\ensuremath{\mat{W}_{\mathrm{RF}}}\xspace} 

\newcommand{\channel}{\ensuremath{\mat{H}}\xspace} 

\newcommand{\Hsi}{\ensuremath{\channel_{\mathrm{SI}}}\xspace}
\usepackage{bm}

\usepackage{xspace}
\usepackage[acronym,nogroupskip,nonumberlist,nopostdot]{glossaries}
\makeglossaries

\usepackage{xcolor}

\newacronym{snr}{SNR}{signal-to-noise ratio}
\newacronym{sinr}{SINR}{signal-to-interference-plus-noise ratio}
\newacronym{inr}{INR}{interference-to-noise ratio}
\newacronym{sir}{SIR}{signal-to-interference ratio}
\newacronym{sqr}{SQR}{signal-to-quantization-noise ratio}
\newacronym{sqnr}{SQNR}{signal-to-quantization-plus-noise ratio}
\newacronym{ian}{IAN}{interference as noise}
\newacronym{ber}{BER}{bit error rate}
\newacronym{pn}{PN}{pseudorandom noise}
\newacronym{bfsk}{BFSK}{binary frequency shift keying}
\newacronym{fh}{FH}{frequency-hopped}
\newacronym{fh-bfsk}{FH-BFSK}{frequency-hopped binary frequency shift keying}
\newacronym{crc}{CRC}{cyclic redundancy check}
\newacronym{isi}{ISI}{intersymbol interference}
\newacronym{dsss}{DSSS}{direct-sequence spread spectrum}
\newacronym{ofdm}{OFDM}{orthogonal frequency-division multiplexing}
\newacronym{ofdma}{OFDMA}{orthogonal frequency-division multiple access}
\newacronym{sdr}{SDR}{software-defined radio}
\newacronym{tx}{TX}{transmitter}
\newacronym{rx}{RX}{receiver}
\newacronym{fdd}{FDD}{frequency-division duplexing}
\newacronym{tdd}{TDD}{time-division duplexing}
\newacronym{fdma}{FDMA}{frequency-division multiple access}
\newacronym{tdma}{TDMA}{time-division multiple access}
\newacronym{sdma}{SDMA}{space-division multiple access}
\newacronym[plural=MPCs]{mpc}{MPC}{multipath component}
\newacronym{mui}{MUI}{multi-user interference}

\newacronym{qam}{QAM}{quadrature amplitude modulation}
\newacronym{mqam}{MQAM}{M-ary quadrature amplitude modulation}

\newacronym{ls}{LS}{least-squares}
\newacronym{lms}{LMS}{least mean squares}
\newacronym{rls}{RLS}{recursive least-squares}
\newacronym{rzf}{RZF}{regularized zero-forcing}
\newacronym{mmse}{MMSE}{minimum mean square error}
\newacronym{lmmse}{LMMSE}{linear minimum mean square error}
\newacronym{mse}{MSE}{mean square error}
\newacronym{fft}{FFT}{fast Fourier transform}
\newacronym{dft}{DFT}{discrete Fourier transform}
\newacronym{dtft}{DTFT}{discrete-time Fourier transform}
\newacronym{ctft}{CTFT}{continuous-time Fourier transform}
\newacronym{ml}{ML}{machine learning}
\newacronym[plural=NNs]{nn}{NN}{neural network}
\newacronym[plural=RNNs]{rnn}{RNN}{recurrent neural network}
\newacronym[plural=ADCs]{adc}{ADC}{analog-to-digital converter}
\newacronym[plural=DACs]{dac}{DAC}{digital-to-analog converter}
\newacronym[plural=FPGAs]{fpga}{FPGA}{field-programmable gate array}
\newacronym{evm}{EVM}{error vector magnitude}
\newacronym{enob}{ENOB}{effective number of bits}
\newacronym{zf}{ZF}{zero-forcing}
\newacronym{rv}{r.v.}{random variable}
\newacronym{omp}{OMP}{orthogonal matching pursuit}
\newacronym{svd}{SVD}{singular value decomposition}

\newacronym{sdp}{SDP}{semidefinite programming}
\newacronym{psd}{PSD}{positive semidefinite}
\newacronym{nsd}{NSD}{negative semidefinite}

\newacronym{agc}{AGC}{automatic gain control}
\newacronym{rf}{RF}{radio frequency}
\newacronym{los}{LOS}{line-of-sight}
\newacronym{nlos}{NLOS}{non-line-of-sight}
\newacronym{ple}{PLE}{path loss exponent}
\newacronym[plural=dB,firstplural=decibels (dB)]{db}{dB}{decibel}
\newacronym[plural=dBm,firstplural=decibel milliwatts (dBm)]{dbm}{dBm}{decibel milliwatts}
\newacronym{pa}{PA}{power amplifier}
\newacronym{lna}{LNA}{low noise amplifier}
\newacronym{cw}{CW}{continuous wave}
\newacronym{papr}{PAPR}{peak-to-average power ratio}
\newacronym{usrp}{USRP}{Universal Software Radio Peripheral}
\newacronym{irr}{IRR}{image rejection ratio}
\newacronym{lo}{LO}{local oscillator}
\newacronym{vm}{VM}{vector modulator}
\newacronym{mmwave}{mmWave}{millimeter wave}

\newacronym{ism}{ISM}{industrial, scientific, and medical}
\newacronym{csma}{CSMA}{carrier-sense multiple access}
\newacronym{csmaca}{CSMA/CA}{carrier-sense multiple access with collision avoidance}
\newacronym{csmacd}{CSMA/CD}{carrier-sense multiple access with collision detection}
\newacronym{mac}{MAC}{medium access control}
\newacronym{phy}{PHY}{physical layer}
\newacronym{4g}{4G}{fourth-generation}
\newacronym{lte}{LTE}{Long-Term Evolution}
\newacronym{4glte}{4G LTE}{\gls{4g} \gls{lte}}
\newacronym{5g}{5G}{fifth-generation}
\newacronym{nr}{NR}{New Radio}
\newacronym{5gnr}{5G NR}{5G New Radio}
\newacronym{ieee}{IEEE}{Institute of Electrical and Electronics Engineers}
\newacronym{wifi}{Wi-Fi}{IEEE 802.11}
\newacronym{lan}{LAN}{local area network}
\newacronym{wlan}{WLAN}{wireless local area network}
\newacronym[plural=BSs]{bs}{BS}{base station}
\newacronym[plural=SBSs]{sbs}{SBS}{small-cell base station}
\newacronym[plural=FD-SBSs]{fdsbs}{FD-SBS}{\gls{fd}-enabled \gls{sbs}}
\newacronym[plural=MBSs]{mbs}{MBS}{macrocell base station}
\newacronym[plural=UEs]{ue}{UE}{user equipment}
\newacronym{ul}{UL}{uplink}
\newacronym{dl}{DL}{downlink}
\newacronym{qos}{QoS}{Quality of Service}
\newacronym{fcc}{FCC}{Federal Communications Commission}
\newacronym{iab}{IAB}{integrated access and backhaul}
\newacronym{fab}{FAB}{fixed access and backhaul}
\newacronym{hetnet}{HetNet}{heterogeneous network}

\newacronym{siso}{SISO}{single-input single-output}
\newacronym{mimo}{MIMO}{multiple-input multiple-output}
\newacronym{sumimo}{SU-MIMO}{single-user \gls{mimo}}
\newacronym{mumimo}{MU-MIMO}{multi-user \gls{mimo}}
\newacronym{bf}{BF}{beamforming}
\newacronym{ca}{CA}{constant amplitude}
\newacronym{ula}{ULA}{uniform linear array}
\newacronym{upa}{UPA}{uniform planar array}
\newacronym{aoa}{AoA}{angle of arrival}
\newacronym{aod}{AoD}{angle of departure}
\newacronym{dof}{DoF}{degrees of freedom}
\newacronym{csi}{CSI}{channel state information}
\newacronym{csit}{CSIT}{\gls{csi} at the transmitter}
\newacronym{csir}{CSIR}{\gls{csi} at the receiver}

\newacronym{fd}{FD}{in-band full-duplex}
\newacronym{hd}{HD}{half-duplex}
\newacronym{si}{SI}{self-interference}
\newacronym{sic}{SIC}{self-interference cancellation}
\newacronym{soi}{SoI}{signal of interest}
\newacronym{asic}{A-SIC}{analog \acrlong{sic}}
\newacronym{dsic}{D-SIC}{digital \gls{sic}}
\newacronym{star}{STAR}{simultaneous transmit and receive}
\newacronym{warp}{WARP}{Wireless Open-Access Research Platform}
\newacronym{bfc}{BFC}{beamforming cancellation}
\newacronym{ipi}{IPI}{inter-panel-interference}
\newacronym{ipic}{IPIC}{inter-panel-interference cancellation}

\newacronym{qcqp}{QCQP}{quadratically-constrained quadratic programming}
\newacronym{cdf}{CDF}{cumulative density function}

\newacronym{elf}{ELF}{extremely low frequency}
\newacronym{slf}{SLF}{super low frequency}
\newacronym{ulf}{ULF}{ultra low frequency}
\newacronym{vlf}{VLF}{very low frequency}
\newacronym{lf}{LF}{low frequency}
\newacronym{mf}{MF}{medium frequency}
\newacronym{hf}{HF}{high frequency}
\newacronym{vhf}{VHF}{very high frequency}
\newacronym{uhf}{UHF}{ultra high frequency}
\newacronym{shf}{SHF}{super high frequency}
\newacronym{ehf}{EHF}{extremely high frequency}
\newacronym{thf}{THF}{tremendously high frequency}

\newacronym{wncg}{WNCG}{Wireless Networking and Communications Group}
\newacronym{linc}{LINC}{Laboratory of Informatics, Networks, and Communications}
\newacronym{ut}{UT Austin}{The University of Texas at Austin}
\newacronym{uiuc}{UIUC}{University of Illinois at Urbana-Champaign}
\newacronym{usc}{USC}{University of Southern California}
\newacronym{mit}{MIT}{Massachusetts Institute of Technology}
\newacronym{berkeley}{UC Berkeley}{University of California, Berkeley}
\newacronym{osu}{OSU}{Ohio State University}

\newcommand{\mmwave}{\gls{mmwave}\xspace}
\newcommand{\mimo}{\acrshort{mimo}\xspace}
\newcommand{\mmse}{\acrshort{mmse}\xspace}

\newcommand{\rf}{\acrshort{rf}\xspace}

\newcommand{\fg}{\gls{5g}\xspace}

\newcommand{\sic}{\gls{sic}\xspace}

\newcommand{\lna}{\acrshort{lna}\xspace}

\newcommand{\lnas}{\acrshortpl{lna}\xspace}
\newcommand{\pas}{\acrshortpl{pa}\xspace}

\newcommand{\adc}{\acrshort{adc}\xspace}
\newcommand{\adcs}{\acrshortpl{adc}\xspace}

\newcommand{\dacs}{\acrshortpl{dac}\xspace}

\newcommand{\snr}{\acrshort{snr}\xspace}

\newcommand{\figref}[1]{\figurename~\ref{#1}}

\begin{document}

%
\title{Millimeter Wave Full-Duplex Radios:\\New Challenges and Techniques}
%
%
%




\author{%
	Ian~P.~Roberts,~%
	Jeffrey~G.~Andrews,~%
	Hardik~B.~Jain,~%
	and Sriram~Vishwanath%
	\thanks{Ian~P.~Roberts and Jeffrey~G.~Andrews are with the Wireless Networking and Communications Group at the University of Texas at Austin. Hardik~B.~Jain and Sriram~Vishwanath are with GenXComm, Inc.  Ian~P.~Roberts was with GenXComm, Inc.~at the time of writing this article.}%
	\thanks{Corresponding author: Ian~P.~Roberts (ipr@utexas.edu).}%
}


%
%

\markboth{}%
 {Roberts \MakeLowercase{\textit{et al.}}: Millimeter-Wave Full-Duplex}
%



\maketitle

\begin{abstract}


Equipping \mmwave systems with full-duplex capability would accelerate and transform next-generation wireless applications and forge a path for new ones. 
Full-duplex \mmwave transceivers could capitalize on the already attractive features of \mmwave communication by supplying spectral efficiency gains and latency improvements while also affording future networks with deployment solutions in the form of interference management and wireless backhaul. 
Foreseeable challenges and obstacles in making \mmwave full-duplex a reality are presented in this article along with noteworthy unknowns warranting further investigation.
With these novelties of \mmwave full-duplex in mind, we lay out potential solutions---beyond active self-interference cancellation---that harness the spatial degrees of freedom bestowed by dense antenna arrays to enable simultaneous transmission and reception in-band.

 
\end{abstract}

\glsresetall

\section*{Introduction} \label{sec:introduction}


New communication systems like \fg cellular and IEEE 802.11ad/ay harness the wide bandwidths available at \mmwave frequencies (roughly 30--100 GHz) to meet the ever-growing demand for high-rate wireless access \cite{rappaport_millimeter_2013}. 
Cellular and local area \mmwave communication systems rely on high beamforming gains provided by dense antenna arrays---on the order of dozens or hundreds of elements---to overcome the high path loss at \mmwave frequencies and achieve sufficient link margin. 
Hybrid digital/analog beamforming architectures offer an efficient means to control these dense arrays with a reduced number of \rf chains, making them ubiquitous in practical \mmwave transceivers \cite{heath_overview_2016}.

Concurrent to recent research on \mmwave communication has been the development of in-band full-duplex technology---a long sought after capability that allows a device to simultaneously transmit and receive across the same frequencies \cite{sabharwal_-band_2014}.
Full-duplex has come a long way in the past decade, particularly in sub-6 GHz transceivers, largely thanks to novel and effective active \sic strategies (e.g., analog and digital \sic) that can rid a desired receive signal virtually free of self-interference.

Equipping \mmwave systems with full-duplex would transform what is capable at the physical layer and in medium access in next-generation networks.
Most obviously, the throughput gains provided by full-duplex would be magnified by the wideband, high-rate communication that is inherent to \mmwave systems.
Latency---a driving metric in future \mmwave applications and networks---is improved with full-duplex since delays associated with half-duplexing transmission and reception can be avoided. 
The deployment of dense \mmwave networks can be made more cost effective with full-duplex \acrlong{iab} solutions---reducing the required density of fiber connectivity (a key hurdle to \mmwave deployments) and preserving precious spectrum.
Furthermore, in unlicensed \mmwave spectrum (e.g., the 57--64 GHz \acrshort{ism} band) and other lightly used bands, full-duplex can introduce new approaches for coexistence between communication, consumer radar, and other incumbents.

Existing solutions for full-duplex at lower frequencies do not immediately translate to \mmwave systems due to fundamental differences between \mmwave and sub-6 GHz transceivers.
As noted, \mmwave systems utilize many more antennas over much wider bandwidths and have unique transceiver architectures and system-design challenges. 
As we will discuss, analog \sic is not well-suited for the dense arrays and wide bandwidths found in \mmwave communication.
Beyond active cancellation, \mimo precoding and combining strategies were explored in sub-6 GHz full-duplex, which aim to mitigate the self-interference by exploiting spatial degrees of freedom \cite{everett_softnull_2016}.
These \mimo-based approaches offer inspiration for \mmwave full-duplex solutions, but features such as hybrid beamforming, wide bandwidths, high sampling rates, beam alignment, and propagation characteristics will dictate what is possible and practical at \mmwave.
While passive approaches (e.g., highly directive antennas, polarization separation) have been proposed for \mmwave full-duplex, this article pertains to transceivers with dense antenna arrays and assumes the use of passive methods could potentially supplement the discussions herein.

We begin the remainder of this article by highlighting the unique challenges and considerations of \mmwave full-duplex, chiefly hybrid beamforming and the self-interference channel. 
Then, we present several promising directions for achieving \mmwave full-duplex. 
In particular, we will discuss how the spatial domain---which presents some of the key challenges at \mmwave---can in fact be harnessed to enable beamforming-based approaches to mitigating self-interference.
Throughout this article, we aim to inform and inspire readers on the challenges, unknowns, and potential solutions surrounding \mmwave full-duplex, with the hope they identify research problems to pursue.



\section*{The Implications of Hybrid Beamforming Architectures}

\begin{figure*}
	\centering
	\includegraphics[width=\linewidth,height=0.37\textheight,keepaspectratio]{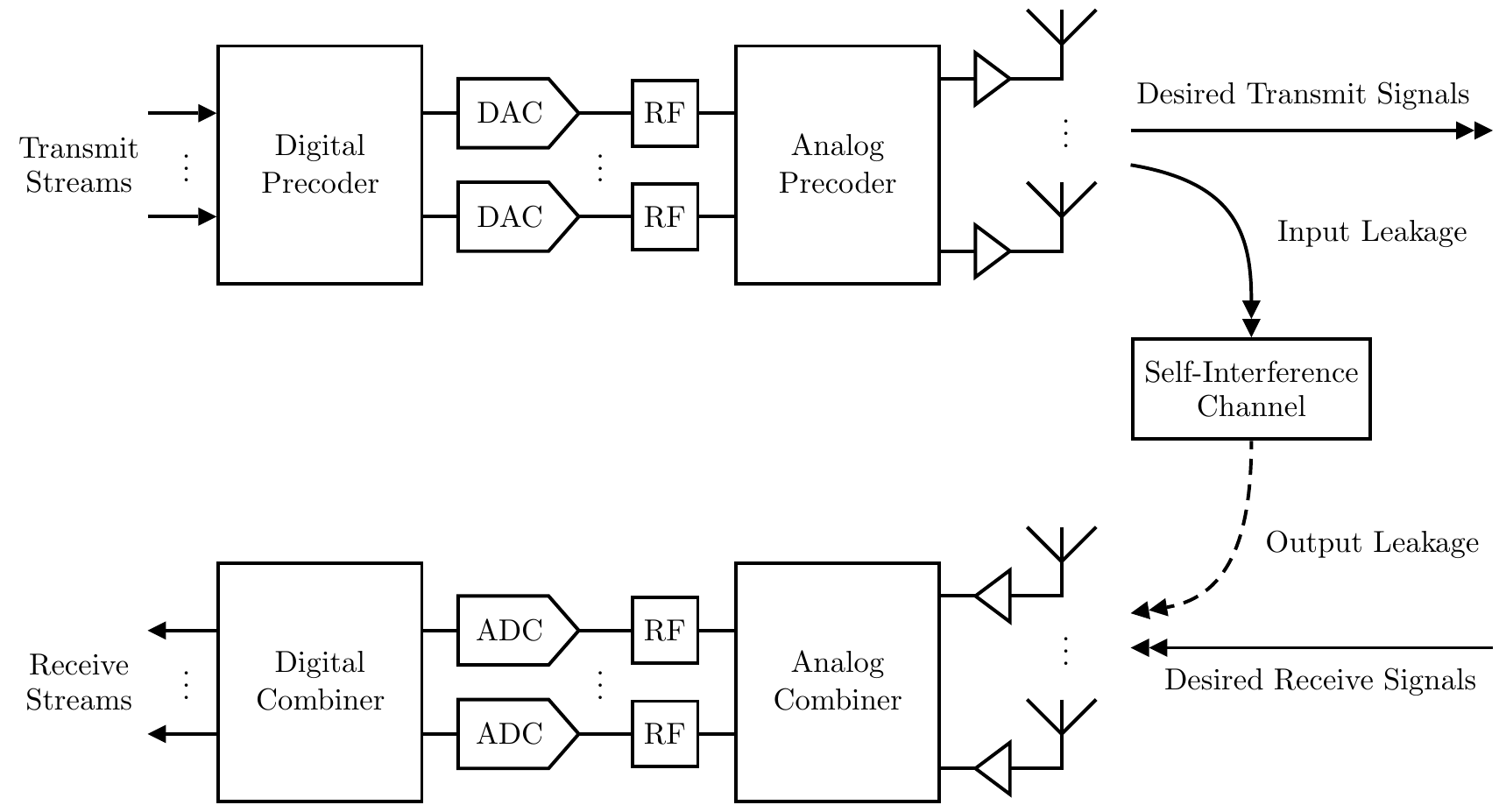}
	\caption{A full-duplex \mmwave transceiver employing hybrid beamforming with separate transmit and receive arrays. Double-ended arrows indicate multi-antenna signals.}
	\label{fig:fdx}
\end{figure*}

A plausible full-duplex \mmwave transceiver is depicted in \figref{fig:fdx}, where separate, independently controlled arrays are used for transmission and reception.
The use of separate arrays appears to be the most practical approach, given that wideband, small form-factor circulators with sufficient isolation for \mmwave full-duplex are still out of reach \cite{dinc_2017}.
A full-duplex \mmwave transceiver aims to transmit to a distant receiver while receiving from a distant transmitter in-band.
The self-interference channel between the transmit array and receive array of a full-duplex \mmwave transceiver is discussed in the next section.
In this section, key system-level implications of \mmwave transceiver architectures on full-duplex are outlined; 
a more detailed, component-wise analysis would be valuable future work.

For \mmwave communication, planar arrays on the order of 16--256 antenna elements are typical. 
To operate these dense antenna arrays with a reduced number of \rf chains, \mmwave transceivers often employ the combination of analog and digital beamforming, termed ``hybrid beamforming'', which provides high beamforming gain while also supporting spatial multiplexing.
Digital beamforming takes place at baseband in software or digital logic, whereas analog beamforming is implemented at passband (i.e., \rf) in analog as a network of phase shifters and possibly attenuators.
The phase shifters and attenuators making up a practical analog beamformer are likely controlled digitally, subjecting them to phase and amplitude quantization, respectively.
Fortunately, highly focused beams can be constructed with low-resolution phase shifters and attenuators.
The ability to create arbitrary beams, however, is lost with quantized phase and amplitude control, which may restrict beamforming-based full-duplex solutions.
Furthermore, the mathematical constraints imposed by quantized phase and amplitude control (chiefly their non-convexity) complicate the optimization of beamforming-based solutions for full-duplex, though efficient solutions to overcome these complications in half-duplex settings may offer inspiration \cite{alkhateeb_channel_2014,sohrabi_hybrid_2015}.
While most \mmwave research ignores amplitude control in analog beamforming for half-duplex settings, we expect it will be extremely useful in tailoring beamforming-based solutions for \mmwave full-duplex.

Given that communication at \mmwave harnesses channels on the order of hundreds of megahertz or even gigahertz, frequency-selectivity is a serious concern for both beamforming-based and filter-based full-duplex solutions.
Even with OFDM, frequency-selective beamforming is not straightforward with hybrid beamforming.
Unlike digital beamforming, which can beamform on a per-subcarrier basis, analog beamforming is (relatively) frequency-flat, treating all subcarriers equally. 
Filter-based approaches for \mmwave full-duplex will require several taps to address frequency-selectivity and will need to have wideband support.

Notice that \figref{fig:fdx} depicts the \pas and \lnas as placed per-antenna, as is common to avoid losses and noise before and after the antenna, respectively.
The \dacs and \adcs, along with upconversion and downconversion, exist in the \rf chains of the transceiver.
The placement of these components is of immense importance to full-duplex systems needing to address nonlinearity and receiver-side saturation \cite{korpi_full-duplex_2014}.
The several high-rate \dacs and \adcs, along with the dozens or hundreds of \pas and \lnas, will play a critical role in the requirements and design of \mmwave full-duplex solutions.
Practical implementations may resort to more affordable, lower quality components---particularly the numerous \pas and \lnas---which may complicate full-duplex.
Generally, wideband \mmwave communication takes place in low-\snr regimes, largely due to propagation losses and a high integrated noise power.
A raised noise floor actually relaxes full-duplex requirements to a degree since more self-interference can be tolerated while remaining noise-limited.
Furthermore, low-\snr communication demands fewer bits of quantization, meaning more self-interference can be tolerated at the \adcs (relative to the desired signal) or, alternatively, lower-resolution \adcs can be used, which is particularly attractive given their high sampling rate requirements.


\section*{Self-Interference Channels at mmWave: How to Model and Estimate Them?}

When attempting to transmit and receive from a \mmwave full-duplex transceiver, self-interference is inflicted by each transmit element onto the entire receive array, collectively producing a large \mimo channel (e.g., of size 64$\times$64).
This high-dimensional over-the-air self-interference channel is an important difference between \mmwave and conventional sub-6 GHz full-duplex and motivates many points of discussion throughout this article.
As shown in \figref{fig:fdx}, we refer to the portion of the transmit signals that get transformed by the self-interference channel as ``input leakage'' and the resulting interference striking the receive array as ``output leakage''. 

At this time, we are not aware of existing work on modeling or measuring the self-interference channel at \mmwave frequencies.
A reasonable starting point has been presented in \cite{xia_2017}, which we outline as follows.
A consequence of the close-in nature of a full-duplex \mmwave transceiver's arrays is that their separation likely does not meet the far-field condition (e.g., $2 D^2 / \lambda$).
Thus, it is reasonable to assume that its arrays will interact in a near-field fashion to some degree.
Along with near-field effects, reflections off the environment---presumably in the far-field---will inflict additional self-interference.
Combining these two contributors, the \mimo self-interference channel matrix at a given instant can be written in the following manner.
\begin{align} \label{eq:si-channel-rice}
	\Hsi &= G_{\mathrm{SI}} \cdot \left( \underbrace{\sqrt{\frac{\kappa}{\kappa + 1}} \channel^{\mathrm{NF}}_{\mathrm{SI}}}_{\mathrm{near-field}} + \underbrace{\sqrt{\frac{1}{\kappa + 1}} \channel^{\mathrm{FF}}_{\mathrm{SI}}}_{\mathrm{far-field}} \right)
\end{align}
The component $\channel_{\mathrm{SI}}^{\mathrm{NF}}$ captures near-field contributions directly from the transmit array to the receive array (e.g., \cite{spherical_2005}), whereas $\channel_{\mathrm{SI}}^{\mathrm{FF}}$ captures far-field contributions from a reflective environment (e.g., a ray-based model). 
When $\channel_{\mathrm{SI}}^{\mathrm{NF}}$ and $\channel_{\mathrm{SI}}^{\mathrm{FF}}$ are normalized to equal energy levels, the Rician factor $\kappa$ throttles the large-scale power disparity between the two.
The large-scale gain $G_{\mathrm{SI}}$ captures the \rf isolation between the arrays.
It is important to note that the high path loss and penetration loss faced at \mmwave frequencies is helpful in mitigating self-interference, especially the portion stemming from far-field reflections.
While this model has yet to be verified with measurements or electromagnetic simulation software---which are important next steps---it offers a starting point for early research on \mmwave full-duplex.
The delay spread and coherence time of the self-interference channel are difficult to speculate, though they will certainly be pertinent to realizing \mmwave full-duplex.

Practically, channel estimation is difficult at \mmwave for several reasons.
Many strategies have been developed to overcome these challenges, often employing compressed sensing strategies which leverage the spatial and temporal sparsity exhibited by point-to-point \mmwave channels \cite{alkhateeb_channel_2014,heath_overview_2016}.
This sparsity is known to pertain to point-to-point far-field \mmwave channels---such as those between devices in cellular and local area networks---but has not been confirmed to exist in \mmwave self-interference channels.
Thus, existing channel estimation strategies do not necessarily readily apply to the self-interference channel. 
A \mmwave full-duplex solution, therefore, may warrant novel self-interference channel estimation strategies, making it an attractive topic for future work.
These new estimation strategies should aim to have low overhead, given the self-interference channel's size, and perhaps the transmit channel and self-interference channel can be estimated using the same time-frequency resources to further reduce this overhead.
However, characterization and modeling of the self-interference channel will be essential first steps before developing means to estimate it.
For instance, if the self-interference channel is near-field dominant (i.e., $\kappa$ is very large), its estimation and how frequently it is estimated will not be at the hand of the dynamics of the far-field environment (e.g., cars, people).
In such a case, perhaps reliable near-field channel models and/or proper calibration accounting for nearby infrastructure could accelerate or potentially replace self-interference channel estimation.



\section*{Can We Extend Analog and Digital Self-Interference\\Cancellation to mmWave?}
We now turn our attention to potential approaches for enabling \mmwave full-duplex and begin by considering popular full-duplex solutions for sub-6 GHz systems: analog \sic and digital \sic.
Digital \sic aims to mitigate residual self-interference---often both linear and significant nonlinear terms---after analog \sic.
Fortunately, digital \sic remains a promising candidate for \mmwave full-duplex since the number of \rf chains remains low with hybrid beamforming,
though exaggerated impairments in \mmwave transceivers may drive up computational costs if not dealt with beforehand. 
A deeper investigation into digital \sic for \mmwave full-duplex would likely yield many useful insights, particularly on transceiver nonlinearity and other impairments such as I/Q imbalance, carrier frequency offset, and phase noise making it a good topic for future work.

Analog \sic traditionally serves two main purposes in full-duplex: (i) prevent \lna and \adc saturation and (ii) capture and cancel nonidealities introduced by the transmit chain.
To consider analog \sic as a potential full-duplex solution for \mmwave systems, it is important to examine the placement of \mmwave transceiver components, chiefly \pas, \lnas, and \adcs.
It is often the case that \pas and \lnas are per-antenna, meaning they are after analog precoding and before analog combining, respectively, whereas \adcs are per-\rf chain after analog combining.
For an analog \sic solution to capture nonlinear terms introduced by the transmit \pas---which have shown to be a bottleneck in sub-6 GHz full-duplex systems \cite{korpi_full-duplex_2014}---it will need to grow in one dimension with the number of transmit antennas.
Likewise, to prevent \lna saturation, analog \sic will need to grow in its second dimension with the number of receive antennas.
Its third dimension, the delay dimension (number of taps), will be based on the impulse response of the self-interference channel.
Considering that \mmwave systems will operate using dozens or hundreds of antennas over wide bandwidths, analog \sic solutions would presumably be relatively large in all three dimensions at \mmwave and, as a result, likely prohibitive in size and complexity.
The obstacles faced by extending analog \sic to \mmwave systems motivate new approaches that address \lna and \adc saturation during full-duplex operation.

\section*{Beamforming Cancellation: Giving \mmwave Full-Duplex Some Space}
The dense antenna arrays at \mmwave seem to complicate the extension of analog \sic solutions but simultaneously promote the spatial domain as a promising arena for mitigating self-interference.
By strategically transmitting and receiving from its numerous antennas, a \mmwave full-duplex transceiver can potentially mitigate self-interference through various forms of ``beamforming cancellation'' \cite{xia_2017,liu_beamforming_2016,satyanarayana_hybrid_2019,roberts_beamforming_2019,zhu_uav_joint_2020}.
Transmit-side beamforming cancellation aims to reduce the output leakage reaching the receive array while still transmitting to a distant receiver.
Similarly, receive-side beamforming cancellation aims to reject the output leakage while receiving from a distant transmitter.

The principle of beamforming cancellation is to tailor the analog and digital precoders ($\prerf$ and $\prebb$) and the analog and digital combiners ($\comrf$ and $\combb$) of a full-duplex \mmwave transceiver to reduce the strength of the following effective self-interference channel.
\begin{align} \label{eq:bfc}
\underbrace{\combb\ctrans \comrf\ctrans}_{\mathrm{combiners}} \Hsi \underbrace{\prerf \prebb}_{\mathrm{precoders}}  
\end{align}
By reducing the power of the effective self-interference channel through spatial techniques, beamforming cancellation can reduce the presence of self-interference in the time-frequency domain to facilitate simultaneous transmission and reception in-band.
Beamforming cancellation, however, will require more than merely extending existing interference-related \mimo designs due to complications related to hybrid beamforming, the unique intertwining of the self-interference channel, and the need to prevent \lna and \adc saturation.

We envision full-duplex \mmwave transceivers will likely depend on digital \sic to some degree.
Thus, acting as a substitute for analog \sic, the primary objective of beamforming cancellation is to prevent \lna and \adc saturation to preserve the quality of the desired receive signal and to give digital \sic a fighting chance. 
Perhaps, however, much more mitigation can be provided beyond merely addressing receiver-side saturation.

Referring to \figref{fig:fdx}, the effective self-interference channel from the transmitter to per-antenna \lnas is $\Hsi \prerf \prebb$, indicating that the responsibility of preventing \lna saturation lay solely at the transmitter.
Note that transmit power control can always ensure \lna saturation is avoided and would be an attractive tool in conjunction with steering strategies.
On the other hand, the effective self-interference channel from the transmitter to per-\rf chain \adcs is $\comrf\ctrans \Hsi \prerf \prebb$, which indicates that the analog combiner at the receiver can aid the transmitter in preventing \adc saturation.
This is convenient since \adc saturation requirements are often stricter than those of \lna saturation.
The role of the baseband combiner $\combb$ is somewhat arbitrary since it lives in the digital domain---after the \adcs---meaning linear interference rejection can be applied and/or more sophisticated digital \sic algorithms.



\begin{figure}
	\centering
	\subfloat[Transmit-side beamforming cancellation.]{\includegraphics[width=0.5\linewidth]{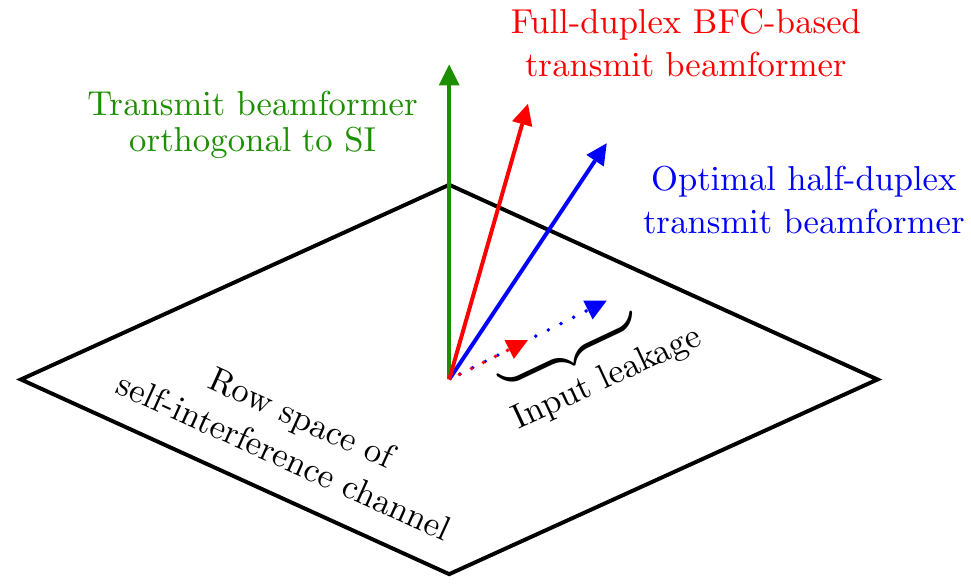}
		\label{fig:vectors-1}}
	\\
	\subfloat[Receive-side beamforming cancellation with user selection.]{\includegraphics[width=0.62\linewidth]{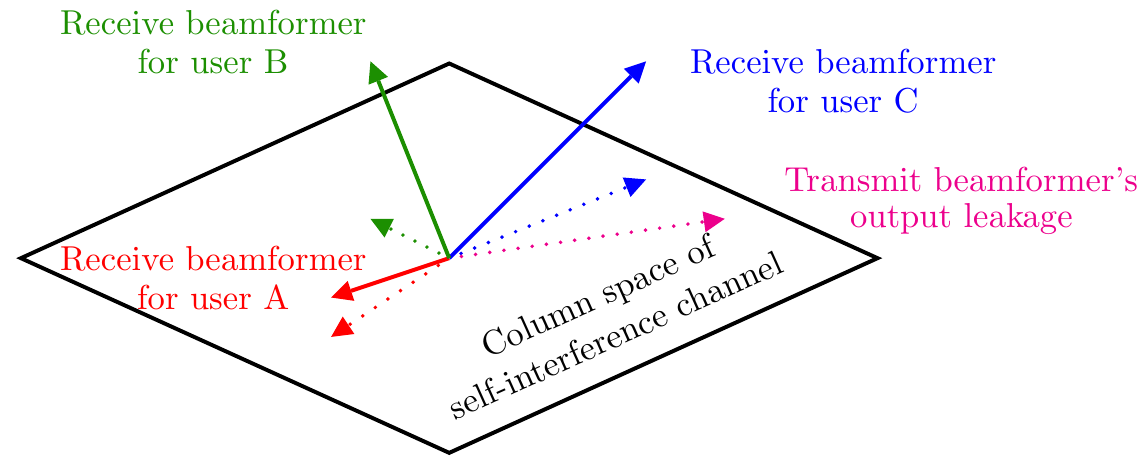}
		\label{fig:vectors-2}}
	\caption{Illustrations of beamforming cancellation.}
	\label{fig:vectors}
\end{figure}

To illustrate a potential beamforming cancellation strategy resembling conventional \acrlong{zf} and \mmse approaches \cite{heath_lozano}, let us consider \figref{fig:vectors-1}.
Depicted is the row space of the self-interference channel matrix $\Hsi$ along with three beamforming vectors. 
The green vector orthogonal to the row space represents a transmit beamformer (i.e., a column of $\prerf$ or $\prerf\prebb$) that would \textit{completely} avoid inflicting self-interference onto the receive array of the full-duplex transceiver (i.e., a zero-forcing approach), driving \eqref{eq:bfc} to zero but perhaps transmitting poorly to a distant user.
The optimal half-duplex transmit beamformer in blue, on the other hand, has a significant component in the row space, potentially leading to a high degree of self-interference onto the receive array.
Between these two beamformers, a potential beamforming cancellation-based transmitter may live, inflicting a reduced amount of self-interference onto the receive array while still transmitting effectively to a distant receiver.
Beamforming at the full-duplex receiver could operate similarly to further reject self-interference. 
We remark that this sort of orthogonality-based approach is merely one of countless beamforming cancellation approaches that could be considered---the exact method employed would depend on one's objective (e.g., weighted sum spectral efficiency, weighted MSE) among other factors.
We further remark that the users being served by the full-duplex device can (and likely should) adjust their transmit and receive strategies in accordance with beamforming cancellation taking place at the full-duplex device.

The degree to which a full-duplex device inflicts self-interference depends not only on its  transmit beamformer but also its receive beamformer as evidenced by \eqref{eq:bfc}.
This intertwining of transmit and receive beamforming can seriously complicate their optimization.
In the illustration of \figref{fig:vectors-1}, we have focused on tackling the input leakage and not the output leakage for this precise reason. 
While the true significance of transmit beamforming cancellation is actually in tailoring the output leakage, by shrinking the strength (i.e., norm) of the input leakage (and not changing its direction), we directly reduce the strength of the output leakage.
Adjusting both the strength and direction of the output leakage would certainly be preferred, but such a strategy demands a joint design of the transmit and receive beamformers.
Like in \figref{fig:vectors-2}, fixing the transmitter thereby collapsing the effective self-interference channel down to the resulting output leakage can be a powerful tool since the number of receive antennas is much greater than the number of transmit streams or even transmit \rf chains.
To summarize, beamforming cancellation is not about orthogonalizing transmission and reception---but rather the output leakage and reception---since transmissions are transformed by the self-interference channel before reaching the receive array. 

\section*{The Costs and Limitations of Beamforming Cancellation}

\begin{figure}
	\centering
	\includegraphics[width=0.95\linewidth,height=0.4\textheight,keepaspectratio]{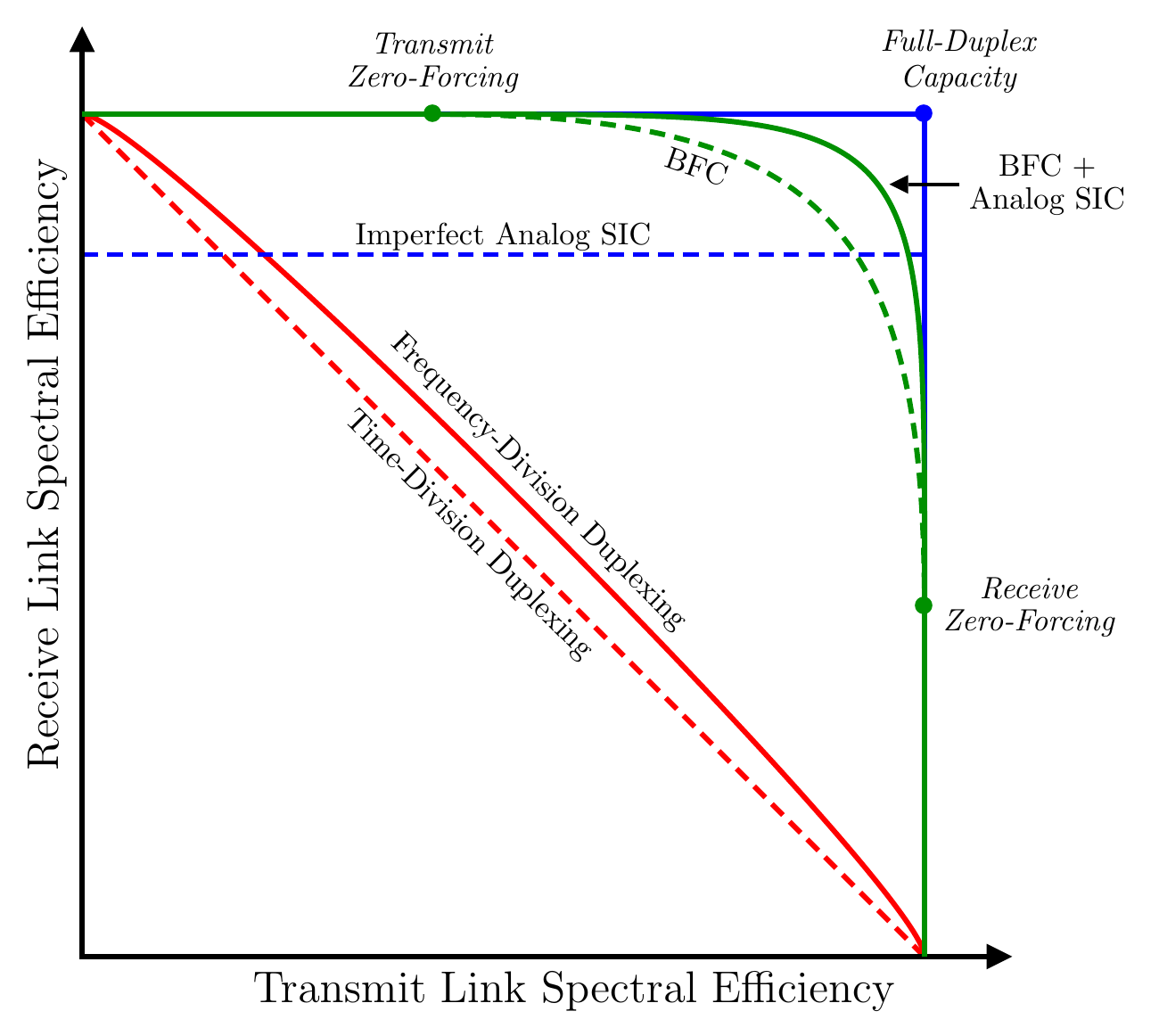}
	\caption{A depiction of the spectral efficiency region boundaries for various duplexing strategies employed by a \mmwave transceiver (not to scale; beamforming cancellation shown as ``BFC''). 
		Supplementing any of these strategies with digital \sic can reduce the gap with the full-duplex capacity.}
	\label{fig:regions}
\end{figure}

Beamforming cancellation is an attractive approach towards \mmwave full-duplex, but it does generally incur some loss in spectral efficiency when compared to the full-duplex capacity\footnote{We define the full-duplex capacity as the maximum sum spectral efficiency afforded by the time-frequency-space resources of the transmit and receive channels.} as it attempts to reduce self-interference spatially. 
Fortunately, the high spatial degrees of freedom provided by \mmwave arrays suggests that this loss is often tolerable since only a few degrees of freedom are used to communicate information, leaving many to tackle self-interference.

\figref{fig:regions} illustrates the spectral efficiency region boundaries of a \mmwave full-duplex transceiver's transmit and receive links.
Perfect execution of analog \sic cancels self-interference completely, achieving the full-duplex capacity since no time-frequency-space resources are consumed to duplex transmission and reception.
Imperfect analog \sic plagues the receive link with residual self-interference, degrading its spectral efficiency.
Beamforming cancellation can approach the full-duplex capacity, though falls short for two reasons:
\begin{enumerate}
\item by deviating from optimal half-duplex strategies on the transmit and receive links, effectively consuming spatial resources
\item by permitting some residual self-interference to more optimally transmit and receive
\end{enumerate}
Supplying beamforming cancellation with analog \sic, as we discuss later, can inch the system closer to the full-duplex capacity.
It is important to note that, unlike relying solely on analog \sic, beamforming cancellation affords the system the ability to trade transmit link performance for receive link performance---a very powerful concept, especially when the transmit and receive links are disparate.
While not always possible, transmit zero-forcing and receive zero-forcing correspond to completely mitigating self-interference via beamforming cancellation, though this is almost certainly not sum-rate-optimal.
Instead, tolerating some self-interference would likely allow a device to more optimally serve users, especially if digital \sic is aiding beamforming cancellation.


Beamforming cancellation designs will certainly need robustness to self-interference channel estimation errors, especially since their effects will be magnified by the relative strength of self-interference.
In addition, hybrid beamforming and per-antenna transmit power constraints may bottleneck the practical realization of a desired beamforming cancellation design, especially in frequency-selective settings.
Importantly, understanding how beamforming cancellation interacts with nonlinear terms induced by the transmit \pas will shed light on the system design of \mmwave full-duplex.
Investigating the severity of these many factors will be critical to fleshing out beamforming cancellation as a potential \mmwave full-duplex solution.

\section*{Leveraging User Selection with Beamforming Cancellation}


Since successful transmission and reception are based on the channels between the full-duplex transceiver and the distant users it is serving, the effectiveness of beamforming cancellation may be highly subject to the environment and of the users being served.
Users whose channels are aligned with the self-interference channel will make beamforming cancellation more costly since significant deviations from optimal half-duplex strategies will be required to mitigate self-interference.
When a full-duplex transceiver employing beamforming cancellation has the liberty of choosing or scheduling the devices it will communicate with, the principle of user selection becomes a powerful tool for interference reduction.
In essence, some transmit-receive pairings naturally afford more self-interference mitigation than others. 
However, we make the distinction that the mitigation afforded by two users is not based on \textit{their} relative orthogonality since the full-duplex device's transmit beam and receive beam are coupled by the self-interference channel.

We illustrate the power of user selection by referencing \figref{fig:vectors-2}.
Fixing the transmit beam, for instance, will inflict some degree of output leakage onto the receive array, shown by the magenta dotted arrow.
Candidates to receive from are transmitting users A, B, and C, which is done so by beamforming along their respective solid vector.
The shadow cast by each of these vectors represents the portion that will foster self-interference.
Choosing the vector whose dotted shadow lay most orthogonal to the output leakage striking the receive array---user B in this case---will reject self-interference most naturally, reducing the spatial resources consumed to achieve full-duplex.

Of course, this overly-simplistic scenario and approach are not flawless, particularly in regard to fairness, since a greedy approach to maximize the sum rate would be to continuously transmit and receive to the ``best'' pair of users. 
Furthermore, asymmetric uplink/downlink demands across users and the dynamics of their channels (e.g., due to mobility) will impact how users are selected for service.
Nevertheless, this toy example illustrates how user selection can be leveraged to improve beamforming cancellation, especially as the number of users available for selection grows.

\begin{figure}
	\centering
	\includegraphics[width=\linewidth,height=0.35\textheight,keepaspectratio]{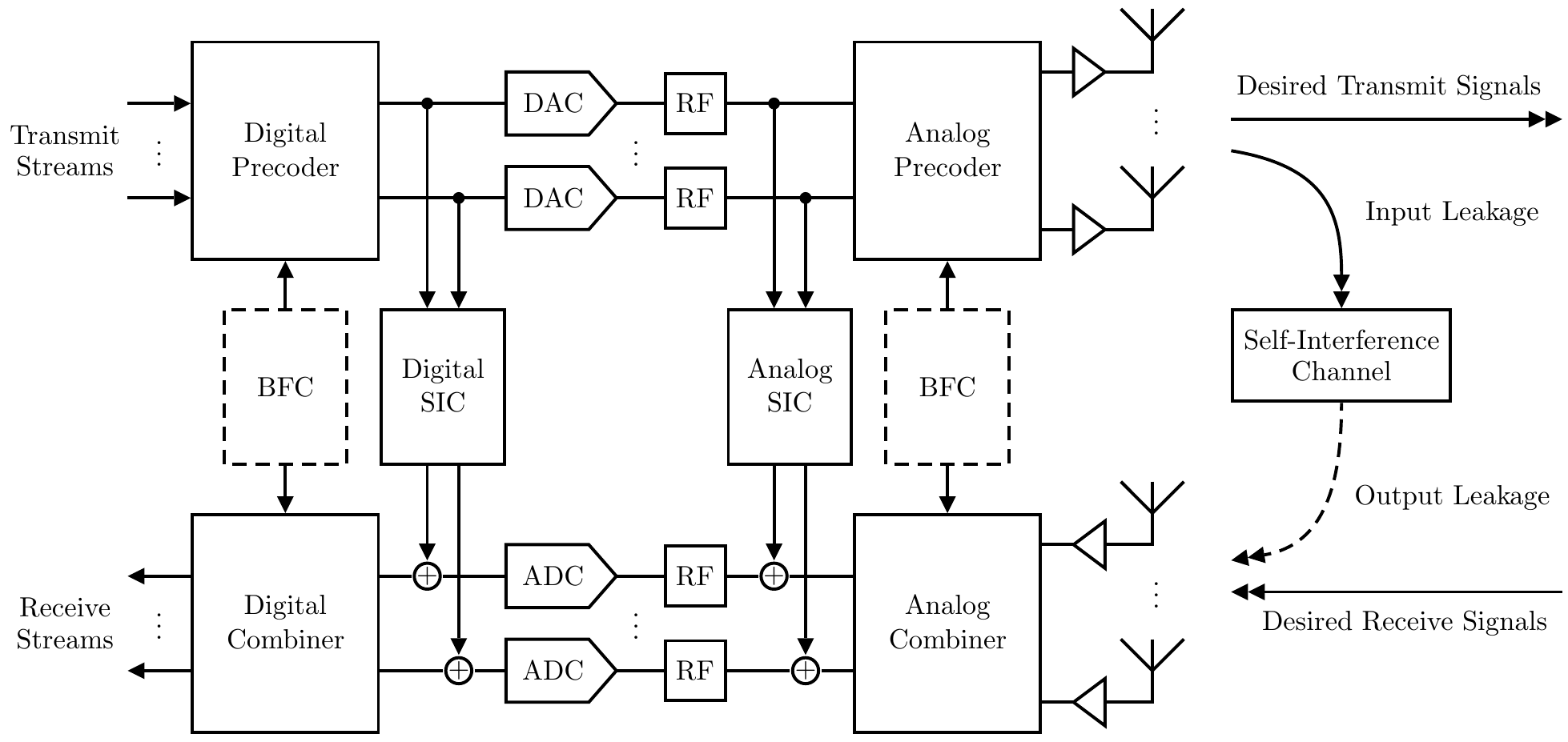}
	\caption{A full-duplex \mmwave transceiver architecture employing beamforming cancellation (shown as ``BFC''), reduced-size analog \sic, and digital \sic.}
	\label{fig:bfc-asic-dsic}
\end{figure}

\begin{figure}[!t]
	\centering
	\includegraphics[width=\linewidth,height=0.5\textheight,keepaspectratio]{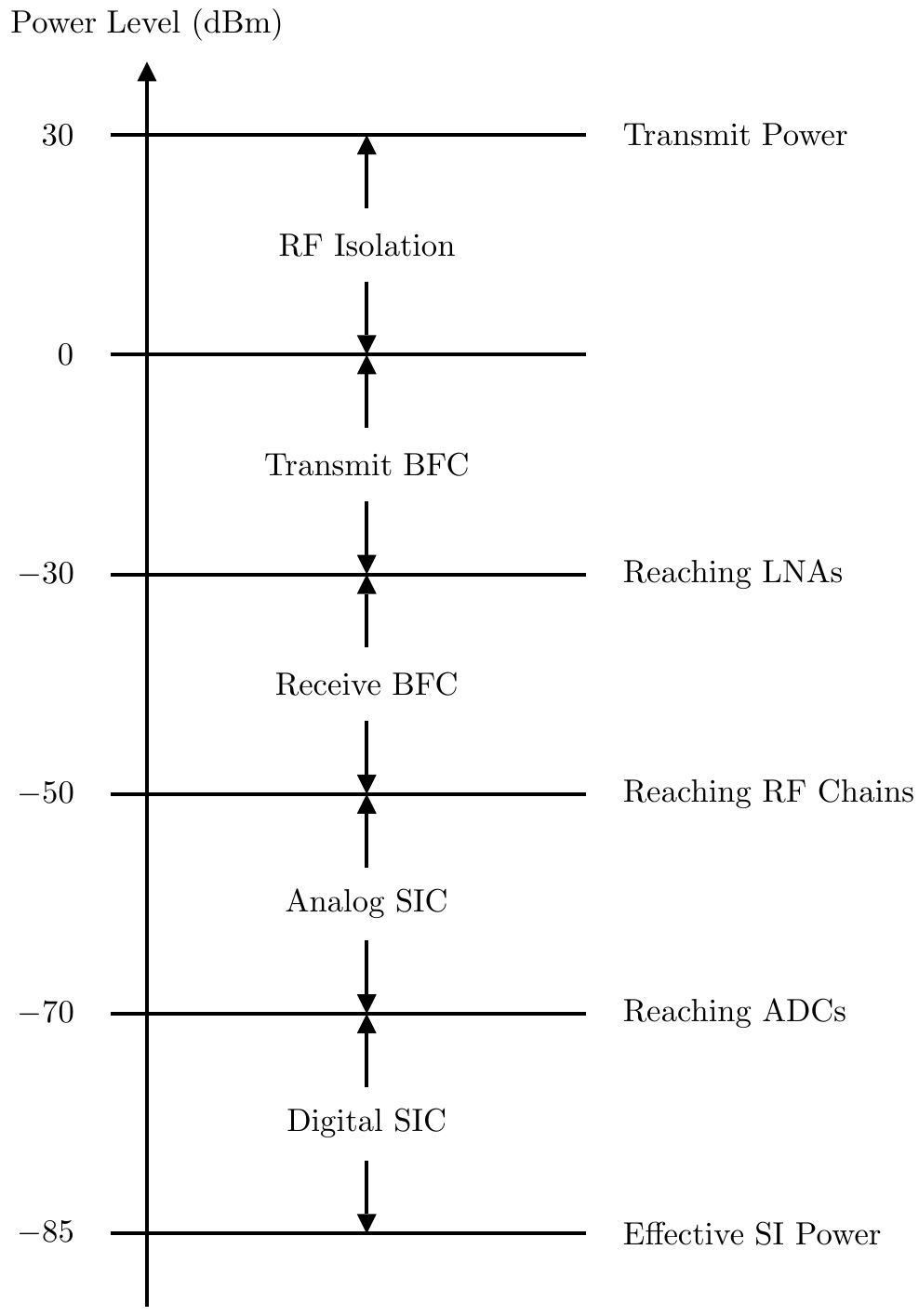}
	\caption{Example breakdown of self-interference power levels at various points in a full-duplex \mmwave transceiver employing beamforming cancellation (shown as ``BFC''), reduced-size analog \sic, and digital \sic.}
	\label{fig:levels}
\end{figure}

\section*{Combining Beamforming Cancellation and Analog\\Self-Interference Cancellation}

While beamforming cancellation could potentially replace the need for analog \sic, we also envision a use for both, considering the potential limitations and performance costs of beamforming cancellation alone.
It may be possible that an analog \sic solution growing with the number of \rf chains---rather than the number of antennas---could aid in achieving \mmwave full-duplex by supplementing beamforming cancellation \cite{roberts_equipping_2020}.
Suppose an analog \sic filter is placed across the transmit and receive \rf chains as shown in \figref{fig:bfc-asic-dsic}. 
It is important to note that, in this case, analog \sic is driven by \rf signals before the analog precoder, meaning it will not incorporate nonlinear terms introduced by per-antenna \pas and other nonidealities.
The advantage of this architecture, however, is that the responsibility of mitigating self-interference is shared across beamforming cancellation, analog \sic, and digital \sic, which we illustrate in \figref{fig:levels}.
Transmit beamforming cancellation would remain \textit{solely} responsible for preventing per-antenna \lna saturation, while analog \sic and beamforming cancellation (transmit and receive side) would share the responsibility of addressing \adc saturation.
Lastly, digital \sic would aim to cancel residual self-interference. 

By taking this approach, analog \sic solutions of reasonable size can aid in mitigating self-interference and preventing receiver-side saturation.
Furthermore, by reducing the responsibility of beamforming cancellation in mitigating self-interference, the full-duplex transceiver can serve users more optimally, as illustrated in \figref{fig:regions}.
In frequency-selective settings, we imagine this staged cancellation approach will be very attractive since it allows both beamforming cancellation and analog \sic to address the selectivity---offering the flexibility to trade frequency-selective beamforming for many-tap analog \sic filters.
Finally, note that analog \sic does not need explicit knowledge of the over-the-air channel; it merely needs to know the significantly reduced effective channel from transmit \rf chains to receive \rf chains (i.e., $\comrf\ctrans \Hsi \prerf$).
This relatively small channel can be observed digitally with conventional estimation strategies---which are likely more reliable and frequent than estimation of the over-the-air counterpart (i.e., $\Hsi$).



\begin{figure}[!t]
	\centering
	\includegraphics[width=\linewidth,height=0.3\textheight,keepaspectratio]{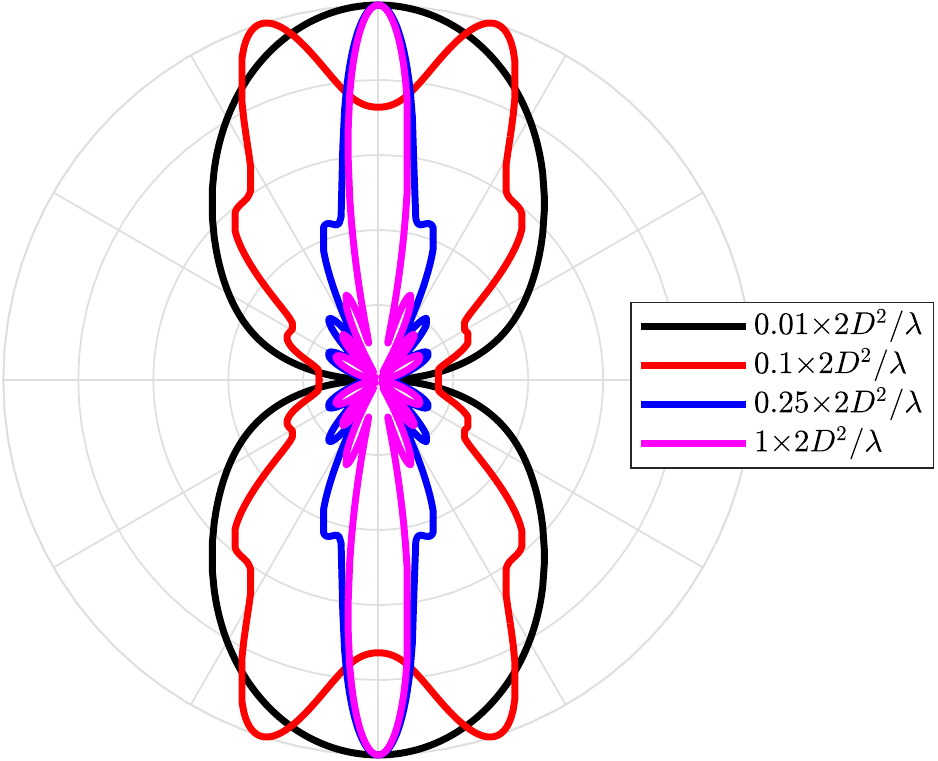}
	\caption{The normalized array factor of an 8-element half-wavelength \acrlong{ula} at fractions of the far-field distance rule-of-thumb.}
	\label{fig:patterns}
\end{figure}

\section*{Custom Analog Beamforming Codebooks for mmWave Full-Duplex}

To address channel estimation and initial access challenges and reduce complexity, practical \mmwave networks have turned to codebook-based beam alignment \cite{junyi_wang_beam_2009}, which is often executed as a search through an analog beamforming codebook for beams that work well (e.g., offer high \snr) between two devices.
Since analog beamforming supplies the high beamforming gain that \mmwave communication relies on, beams in an analog beamforming codebook generally have two properties\footnote{In hierarchical codebooks, we refer to codewords in the finest tier.}: (i) offer high beamforming gain (i.e., are highly directional) and (ii) collectively provide good spatial coverage to ensure a user's location within the service region does not inhibit it from being served.

A codebook with entries satisfying these two properties are generally suitable for half-duplex use-cases but do not necessarily offer much resilience to self-interference.
Extremely narrow transmit beams, for instance, have the potential to inflict substantial self-interference onto the neighboring receive array. 
This is because an ``extremely narrow transmit beam'' is extremely narrow in the far-field and not necessarily so in the near-field.
To illustrate this, see \figref{fig:patterns}, which shows the array factor of an 8-element half-wavelength \acrlong{ula} at various ranges from the center of the array.\footnote{This plot is based on ideal near-field behavior (i.e., \cite{spherical_2005}), free of coupling and other various electromagnetic artifacts, which would further misshape the beams.}
At the far-field distance rule-of-thumb, we get the familiar sinc-like shape with a narrow main lobe in the broadside directions.
Getting closer and closer to the array, the far-field approximation deteriorates and we begin to see the effects of the near-field.
Those exhibited in \figref{fig:patterns} become quite omnidirectional, leading us to see how a near-field self-interference channel can be difficult to avoid with conventional analog beamforming.
Since we expect the self-interference channel at \mmwave may contain a significant near-field portion (not to mention reflections from the environment), we again can see that achieving \mmwave full-duplex is not as simple as merely ``aligning the nulls'' of the transmit and receive beams.

Thus, we anticipate research on custom beamforming codebooks tailored for \mmwave full-duplex to be a promising future direction.
For these custom codebooks to replace those conventionally used for beam alignment, their beams will need to provide sufficient beamforming gain and coverage.
Beams would also ideally reject near-field self-interference as well as any stemming from the far-field, if possible.\footnote{Environmental reflections may be unavoidable with a codebook-based approach given the need for good spatial coverage.}
While potentially a good starting place, it is unlikely that ``off-the-shelf'' codebooks would naturally reject self-interference sufficiently, and the design of a custom codebook with these properties would likely be difficult to perfect.
If the ideal codebook could be designed---offering sufficient isolation between all transmit and receive beam pairs while still providing high gain and adequate coverage---a \mmwave transceiver could  \textit{blindly} operate in a full-duplex fashion with little to no need for supplemental analog or digital \sic.
Given that the design of such a custom codebook depends on the self-interference channel, it may be updated according to the channel dynamics or is created based on the long-term statistics of the channel.
We expect successful custom codebook designs for \mmwave full-duplex to be a fast track to deployment since they could integrate into existing beam alignment schemes and are much simpler to execute (once designed) than the previously described methods.
In addition to beam alignment, custom codebook designs like this could be used to simplify and accelerate general beamforming design and optimization, including that of beamforming cancellation.

\section*{Conclusion: Towards Making mmWave Full-Duplex a Reality} \label{sec:conclusion}

We conclude by summarizing important research directions that will be required to mature \mmwave full-duplex from theory to concept and beyond to practice.
Reliable characterization and modeling of the self-interference channel will provide a foundation on which future research can build.
Following that, self-interference channel estimation strategies can be developed, which may exploit newfound structure or sparsity.
Beamforming cancellation designs subject to the constraints imposed by hybrid beamforming can be explored to better understand what performance guarantees and receive-side saturation requirements can be met in realistic environments.
Investigating how beamforming cancellation can be supplemented by analog \sic will provide rich insights on how the two can jointly tackle self-interference, especially in frequency-selective settings.
A thorough analysis of nonlinear self-interference would facilitate system-level studies involving digital \sic, analog \sic, and beamforming cancellation.
Network-level analyses will indicate the power of user selection and the effects full-duplex has on \mmwave access and backhaul.
Prototyping full-duplex \mmwave systems early on will be essential in identifying unexpected obstacles and steering future research.
Drafting full-duplex-based protocols that integrate well with existing networks will be critical in its standardization and deployment.



\bibliographystyle{bibtex/IEEEtran}
\bibliography{bibtex/IEEEabrv,refs}

%

\section*{Biographies}

\vspace{-2.5cm}

\begin{IEEEbiographynophoto}{{Ian~P.~Roberts}}
	[S'16]~(ipr@utexas.edu)~is a graduate student in the Department of Electrical and Computer Engineering at the University of Texas at Austin, where he is part of the Wireless Networking and Communications Group. 
	He received his B.S.~in electrical engineering from Missouri University of Science and Technology in 2018. 
	He has industry experience developing and prototyping wireless technologies at GenXComm, Inc., Amazon, Sandia National Laboratories, and Dynetics, Inc.
	His research interests are in the theory and implementation of millimeter-wave communication, in-band full-duplex, and communication system optimization.
	He is a National Science Foundation Graduate Research Fellow.
\end{IEEEbiographynophoto}%

\vspace{-2cm}

\begin{IEEEbiographynophoto}{{Jeffrey~G.~Andrews}}
[F'13] is the Cockrell Family Endowed Chair in Engineering at the University of Texas at Austin where his research focuses on wireless communication systems. He received the 2015 Terman Award, the NSF CAREER Award, and the 2019 IEEE Kiyo Tomiyasu technical field award.  Dr. Andrews is an ISI Highly Cited Researcher and has been co-recipient of 15 best paper awards.  He holds a PhD  in Electrical Engineering from Stanford University.
\end{IEEEbiographynophoto}%

\vspace{-2cm}

\begin{IEEEbiographynophoto}{{Hardik~B.~Jain}}
received B.E.~degree in electrical engineering from the Birla Institute of Technology and Science (BITS), Pilani in 2011 and the M.S.~degree in electrical engineering from The University of Texas at Austin in 2015. He is a graduate student currently working towards the Ph.D.~degree in electrical engineering at The University of Texas at Austin. His research interest includes full duplex wireless transmission, design of wireless communication systems and coding theory. His work on full duplex radios and self-interference cancelation technology is being commercialized to build a product at GenXComm Inc. where he serves as the co-founder and CTO. His former industry experience includes work at Texas Instruments, Bangalore in 2011 and Broadcom India Research, Bangalore from 2011--2013.
\end{IEEEbiographynophoto}%

\vspace{-2cm}

\begin{IEEEbiographynophoto}{{Sriram~Vishwanath}}
[SM'08]~received the B. Tech. degree in Electrical Engineering from the Indian Institute of Technology (IIT), Madras, India in 1998, the M.S. degree in Electrical Engineering from California Institute of Technology (Caltech), Pasadena, USA in 1999, and the Ph.D. degree in Electrical Engineering from Stanford University, Stanford, CA, USA in 2003.
Sriram received the NSF CAREER award in 2005 and the ARO Young Investigator Award in 2008. He has published over 250 refereed research papers, across the domains of AI/Machine Learning, Information Theory, Networking, Blockchain and Large Scale Systems.  He is the co-founder of multiple startups in the AI, Networking, Healthcare and Enterprise Software domains.
\end{IEEEbiographynophoto}%

\end{document}